\documentclass[hidelinks,apj]{emulateapj}
\usepackage{gensymb}
\usepackage{subfigure}
\PassOptionsToPackage{hyphens}{url}
\usepackage{hyperref}
\usepackage{lineno}
\usepackage{amsmath}
\newcommand{\Fermi}{{\it Fermi }}
\newcommand{\fermi}{{\it Fermi}}
\newcommand{\FermiL}{{\it Fermi}-LAT }
\newcommand{\fermiL}{{\it Fermi}-LAT}
\newcommand{\Off}{{\it Off }}

\newcommand{\On}{{\it On }}

\usepackage{nicefrac}
\begin{document}

\title{A stacked analysis of 115 pulsars observed by the Fermi LAT} \author{A.~McCann} \affil{The University
  of Chicago, The Kavli Institute for Cosmological Physics, 933 East
  56th Street, Chicago, IL 60637, USA}
\email{mccann@kicp.uchicago.edu}

\begin{abstract}
Due to the low gamma-ray fluxes from pulsars above 50~GeV and the
small collecting area of space-based telescopes, the gamma-ray
emission discovered by the \FermiL in $\sim$150 pulsars is
largely unexplored at these energies. In this regime, the
uncertainties on the spectral data points and/or the constraints from
upper-limits are not sufficient to provide robust tests of competing
emission models in individual pulsars. The discovery of power-law-type
emission from the Crab pulsar at energies exceeding 100~GeV provides a
compelling justification for exploration of other pulsars at these
energies. We applied the method of \textit{Aperture Photometry} to
measure pulsar emission spectra from \FermiL data and present a
stacked analysis of 115 pulsars selected from the Second \FermiL
Catalog of Gamma-ray Pulsars. This analysis, which uses an average of
$\sim$4.2 years of data per pulsar, aggregates low-level emission
which cannot be resolved in individual objects but can be detected in
an ensemble. We find no significant stacked excess at energies above
50~GeV. An upper limit of ∼30\% of the Crab pulsar level is found for
the average flux from 115 pulsars in the 100-177 GeV energy range at
the 95\% confidence level. Stacked searches exclusive to the young
pulsar sample, the millisecond pulsar sample, and several other
promising sub-samples also return no significant excesses above
50~GeV.
\end{abstract}


\section{Introduction}
The Large Area Telescope (LAT) on board NASA's \Fermi satellite has
detected over 150 new gamma-ray
pulsars\footnote{\url{https://confluence.slac.stanford.edu/display/GLAMCOG/Public+List+of+LAT-Detected+Gamma-Ray+Pulsars}}. One
seemingly unifying feature seen in all of these pulsars is the form of
their spectral energy distribution (SED) which is typically described
by a power law followed by a spectral cutoff occurring between 1 and
10~GeV \citep{Abdo2013ApJS}. The relatively narrow range of the
measured cutoff energy observed in pulsars across a wide range of spin
parameters suggests the gamma-ray emission mechanism is common across
these pulsars and that it is not strongly dependent on the pulsar spin
or energetics. Above the GeV break energy, pulsar SEDs appear to fall
exponentially, although for bright gamma-ray pulsars with high
statistics above the break, sub-exponential cutoffs are preferred
\citep{Abdo2013ApJS}.  Curvature radiation occurring at the
radiation-reaction limit in the outer magnetosphere can largely
explain these spectral features and thus, models based on this
mechanism have become the most favored explanations of pulsar emission
in the \FermiL era.

From a modeling perspective, the maximum energy of the observed
radiation provides robust and model-independent constraints on the
altitude of the emission sites above the neutron star
\citep{Story2014ApJ}. The shape of the SED above the break is also a
key probe of the emission mechanisms, with exponential cutoffs
predicted in curvature emission models (e.g.,
\citealt{Harding2008ApJ}) and power-law extensions predicted in
inverse-Compton scattering models (e.g., \citealt{Lyutikov2012ApJb}).
While curvature emission is broadly accepted as the dominant emission
mechanism, many authors, such as \cite{Cheng1986ApJ, Romani1996ApJ,
  Hirotani2001ApJ} and \cite{Takata2007ApJ} have discussed the role of
inverse-Compton emission processes in pulsar magnetospheres. The
discovery of $>$100~GeV emission from the Crab pulsar
\citep{Aliu2011Sci,Aleksi2011ApJ,Aleksi2012A}, which strongly
disfavors curvature-emission-based models, has led to renewed interest
in inverse-Compton studies and \cite{Du2012ApJ} and
\cite{Lyutikov2013MNRAS}, and others, have presented magnetospheric
inverse-Compton models which explain emission above the GeV break
measured by \fermiL. Others, such as \cite{Aharonian2012Natur} and
\cite{Petri2012MNRAS} have presented inverse-Compton models where
pulsed emission at sub-TeV energies originates from acceleration zones
outside the light cylinder.

The Crab pulsar is the only pulsar known which clearly exhibits
non-exponentially-suppressed emission above the GeV break, with
power-law-type emission seen to extend to 400~GeV\footnote{Recently
  the MAGIC collaboration has presented evidence indicating that the
  power-law spectrum of the Crab pulsar may extend to TeV
  energies. See
  \url{http://fermi.gsfc.nasa.gov/science/mtgs/symposia/2014/abstracts/185}}. Studies
of the Geminga pulsar show that the SED above the GeV break is
compatible with a steep power law \citep{Lyutikov2012ApJ,Aliu2014Gem},
but no emission has been seen above 100~GeV. Recent studies of the
Vela pulsar with \FermiL data have reported significant emission above
50 GeV \citep{Leung2014}, but again, no significant emission above
100~GeV has been observed. \cite{Aliu2014Gem} argues that even in the
case of these two bright pulsars - the brightest known to exist - the
available spectral data are not sufficient to discriminate between
exponential and power-law -shaped spectra above $\sim$10~GeV.
The \fermi-LAT catalog of sources above 10~GeV (1FHL) reports the
detection of significant pulsations from thirteen pulsars at energies
which exceed 25~GeV \citep{Ackermann2013ApJS}. These thirteen
pulsars are largely drawn from the brightest pulsars observed by the
\FermiL (F$_{\rm >100~MeV}>1.6\times10^{-7}$s$^{-1}$cm$^{-2}$) and
thus are sufficiently bright to be detected by \FermiL at these
energies even as their spectrum falls rapidly above the break. In the
second \FermiL catalog of gamma-ray pulsars, only four pulsars have a
measured flux point above 30~GeV \citep{Abdo2013ApJS} and the
statistical uncertainties on the spectral points are too large provide
robust tests of model predictions at these energies. The question of
whether the Crab pulsar is unique, or whether
non-exponentially-suppressed gamma-ray spectra are common in gamma-ray
pulsars is of great importance. Beyond pulsar emission modeling,
questions concerning the emission properties of pulsars have
significant implications for galactic dark matter searches, where
unassociated gamma-ray excesses can be interpreted as the remnants of
dark matter annihilation (e.g., \citealt{Abazajian2012PhRvD}). Since
pulsars are likely the main background for these searches,
categorizing the shape of pulsar spectra is a critical step towards
validating any indirect dark matter signal in the gamma-ray domain. To
address these questions, we present a stacked analysis of gamma-ray
pulsars to search for non-exponentially-suppressed emission above
50~GeV which cannot be resolved in the \FermiL analysis of a single
object, but can be detected from an ensemble.

The remainder of this paper is structured in the following way. In
Section 2 we describe the \FermiL data used in this analysis. Section
3 introduces the \textit{Aperture Photometry} method for measuring
pulsar spectra and presents an assessment of its performance. In
Section 4 we present the results of the stacking analysis and in
Section 5 we provide some discussion and concluding remarks.

\section{The Pulsar Data Sample}
\begin{figure}
\centering 
\includegraphics[width=0.49\textwidth]{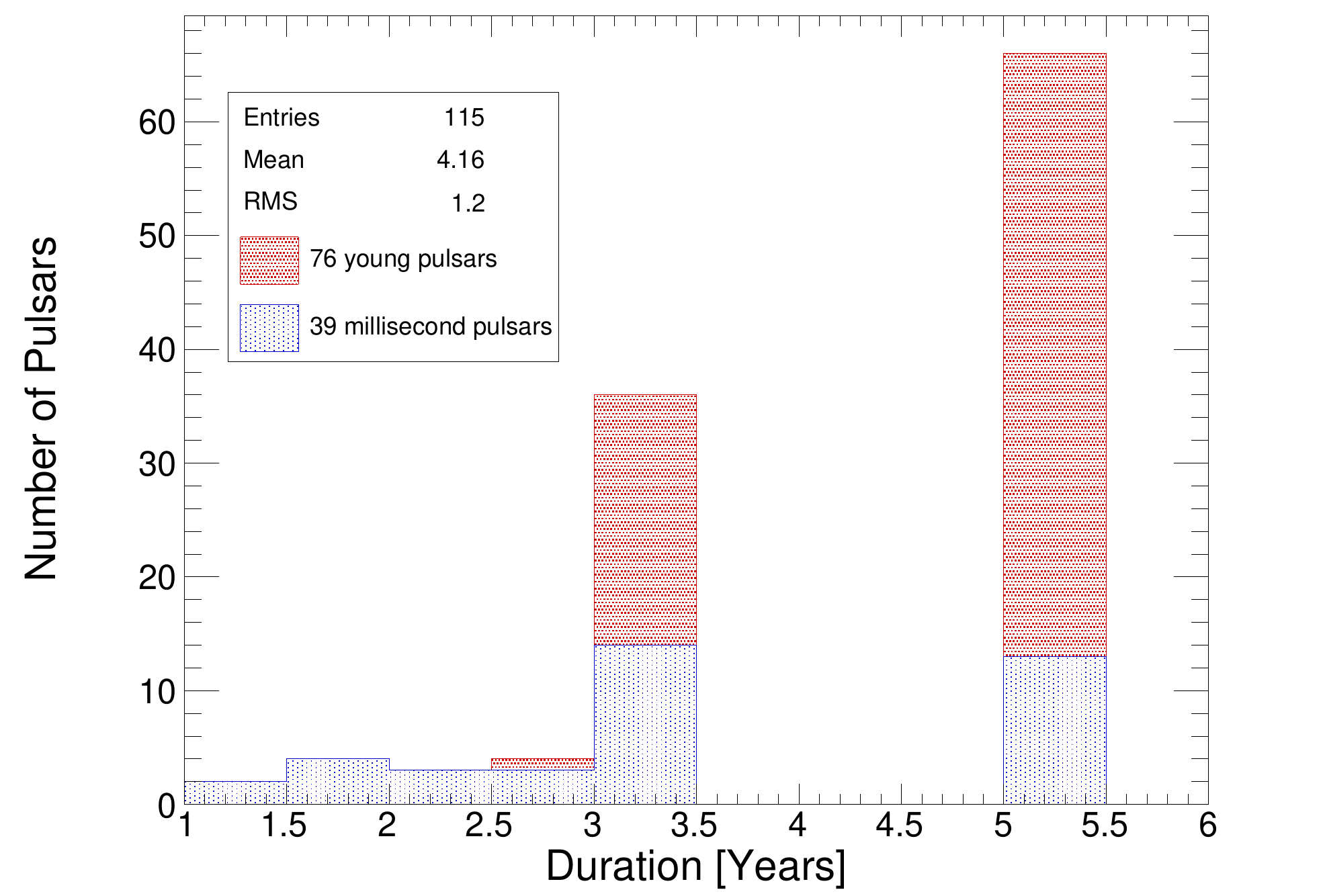}
\caption{\small A histogram of the duration of the individual pulsar data
  sets used in this analysis. The duration is entirely determined by
  the period of validity of the available pulsar timing solutions.}
\label{fig:years}
\end{figure}
This work is based on the 117 pulsars described in the Second \FermiL
Catalog of Gamma-ray Pulsars \citep{Abdo2013ApJS}, which shall be
referred to as 2PC throughout. 2PC provides a wealth of measured
spectral and temporal characteristics which enable this study. Of
particular importance to this analysis is the \Off peak phase range,
which is derived from a Bayesian Block analysis of the pulsar light
curve \citep{Jackson2005,Scargle2013ApJ}. This is the interval of the
pulsar rotation phase where the pulsed emission is deemed to be at its
lowest flux. No \Off range was defined in the 2PC for PSRJ2215+5135,
thus this pulsar was excluded from the study. Further, while used as
the prototype for the analysis methodology, the Crab pulsar was also
excluded from this study, since we are investigating whether
Crab-pulsar-like emission above $\sim$50~GeV is seen in other
pulsars. The exclusion of these two sources reduces the number of
pulsars analyzed in this work to 115 -- 76 young (non-recycled)
pulsars and 39 millisecond pulsars.

Beyond the number of pulsars analyzed, the size of the data set used
for a given pulsar analysis is entirely dependent on the availability
of pulsar timing solutions. Timing solutions encode the spin-down
behavior of the pulsar and enable the conversion of the measured
photon arrival time to the pulsar rotation phase value. Pulsar timing
solutions with periods of validity ranging from 1.4~yr to 3.3~yr were
provided in the supplementary material of the 2PC. For 66 pulsars,
updated timing solutions provided by Matthew
Kerr\footnote{\url{www.slac.stanford.edu/~kerrm/fermi_pulsar_timing/}}
\citep{Kerr2014} which have periods of validity lasting $\sim$5~yr
were chosen over the shorter 2PC timing solutions. Combining the two
cases, the average data set analyzed spans 4.2~yr (see
Figure~\ref{fig:years}). It should be noted, however, that longer
timing solutions were more often provided for brighter pulsars.

The \fermi-LAT pulsar analysis presented here uses Pass-7 reprocessed
\texttt{Source}-class photon data retrieved from the \Fermi science
support center weekly data server. Data processing was performed using
the \Fermi Science Tools version \texttt{v9r33p0-fssc-20140520}.

\section{Aperture Photometry Pulsar Analysis}\label{sec:AP}
\begin{figure*}
\centering 
\includegraphics[width=0.95\textwidth]{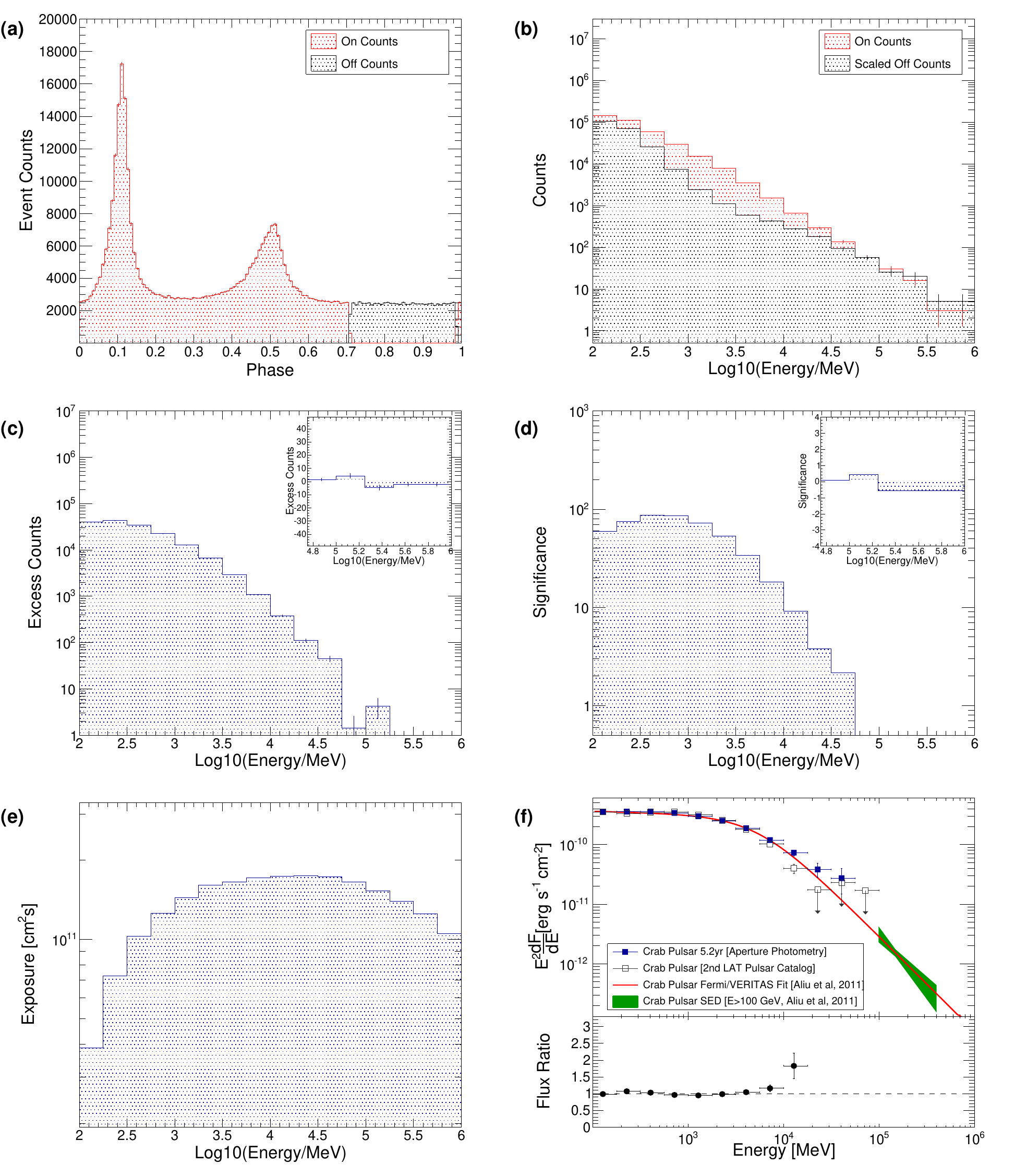}
\caption{\scriptsize Aperture photometry analysis steps for the Crab pulsar. Panel
  (a) plots the phase distribution (light curve) of the Crab pulsar
  from 5.2 years of \fermi-LAT observations. The \Off phase range,
  [0.71 $-$ 0.99], is defined in the 2nd \fermi-LAT catalog of
  gamma-ray pulsars (2PC). Panel (b) plots the distribution of photon
  energies for events which fell in the \On and \Off phase ranges. The
  \Off events have been scaled by $\alpha$ which is the ratio of the
  \On phase gate(s) size to the \Off gate(s) size . Panel (c) shows
  the energy distribution of the excess events and panel (d) shows the
  significance of the excess in each energy bin. Panel (e) shows the
  \fermi-LAT exposure for the ROI used in each energy bin determined
  from \texttt{gtexposure}. In panel (f) the Crab pulsar AP SED is
  plotted alongside the Crab pulsar SED determined from a likelihood
  fit done in the 2PC. A broken power-law fit to \FermiL and VERITAS
  data from \cite{Aliu2011Sci} is plotted, as well as the VERITAS
  $>$100~GeV bow-tie. Below the SED plotted in panel (f) is the ratio
  of the AP flux to the 2PC flux in each bin, showing the level of
  agreement between the AP method and the likelihood method. }
\label{fig:Crab}
\end{figure*}
Spectral analysis of \fermi-LAT data is typically performed through a
maximum likelihood fitting procedure where the photon event data are
fit to a positional and spectral model. The flux for a particular
source is then derived from the best fit model. The \fermi-LAT data
can also be analyzed with an \textit{Aperture Photometry} (AP) method
where the raw event counts from a particular region of interest on the
sky are combined with a measure of the instrument exposure (cm$^{2}$\,s)
to the region to determine the flux. The on-line \fermi-LAT analysis
manual reports that the AP method is less accurate and less sensitive
than the likelihood fitting procedure but that it ``provides a model
independent measure of the flux'' and it ``is less computationally
demanding''\footnote{\url{http://fermi.gsfc.nasa.gov/ssc/data/analysis/scitools/aperture_photometry.html}}. We
demonstrate here that the AP method can be used to produce accurate
SEDs for pulsars from multi-year data sets. This type of analysis is
possible due to the accurate determination of the background rate
which can be measured in the \Off phase range.
\subsection{Aperture Photometry Analysis Steps}\label{sec:APSteps}
\begin{figure*}
\centering
\includegraphics[width=0.99\textwidth]{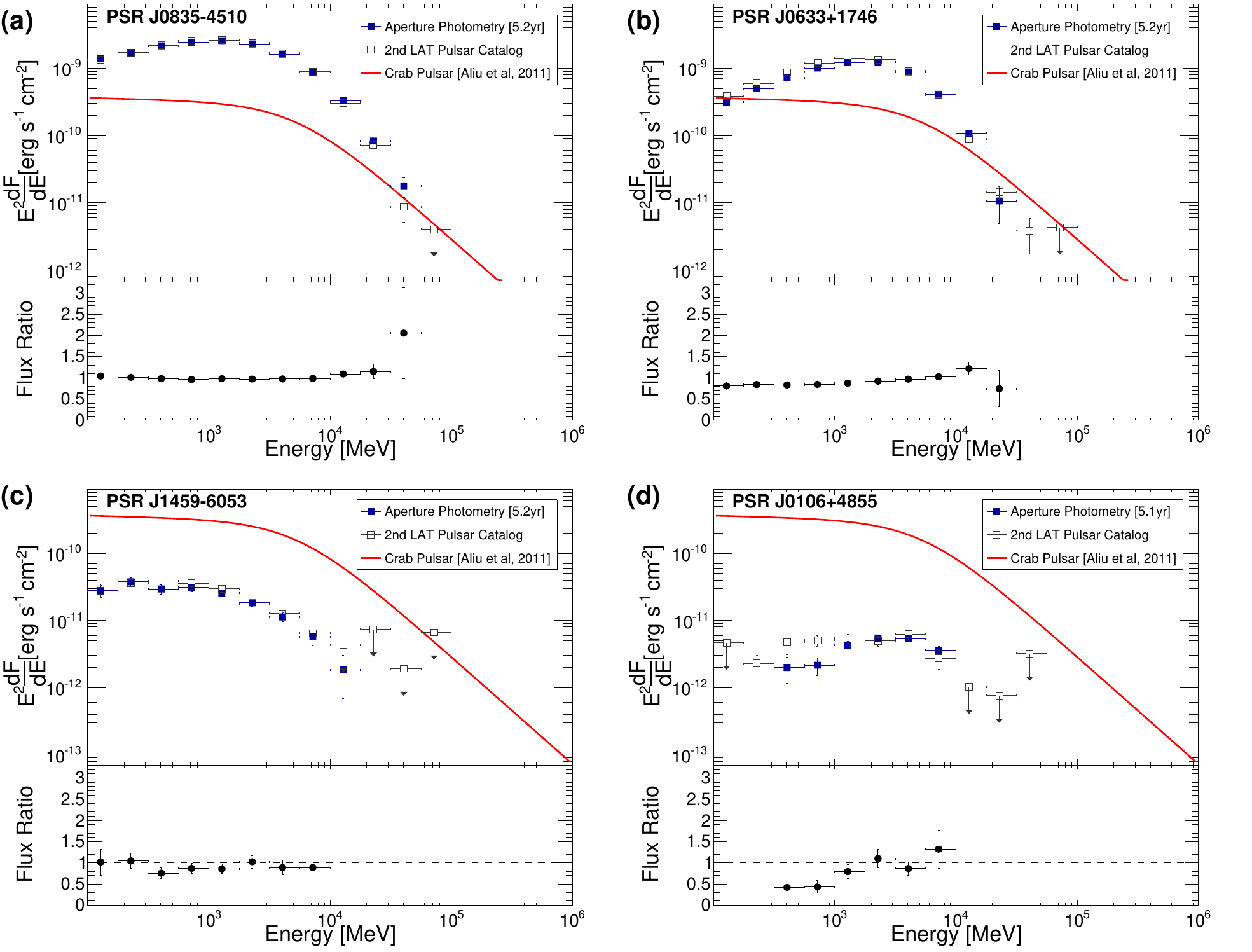}
\caption{\small Example SEDs determined with the AP method. The well known
  Vela and Geminga pulsars are shown in panels (a) and (b)
  respectively, with two newly-discovered dimmer gamma-ray pulsars
  shown in panels (c) and (d). In each panel the AP SED is plotted
  alongside the SED determined from a likelihood fit done in the
  2PC. A broken power-law fit to the Crab pulsar data from
  \cite{Aliu2011Sci} is plotted for scale. Below each SED plot, the
  ratio of the AP flux to the 2PC flux is plotted for bins where a
  $>$1.5$\sigma$ excess is derived in the AP analysis and a flux value
  is reported by the 2PC.}
\label{fig:sed}
\end{figure*}

The processing follows closely the AP analysis thread detailed in the
on-line \fermi-LAT analysis manual\footnotemark[\value{footnote}].
\begin{enumerate}
\item Logarithmically-spaced energy binning with 4 bins per decade is
  chosen over the 100~MeV to 1~TeV energy range.
\item A region of interest (ROI) is chosen around each pulsar with an
  energy-dependent radius. The radius chosen is three times the 68\%
  point-spread-function (PSF) containment radius determined from a
  Vela analysis by \cite{Ackermann2013ApJ} which range from
  2.26$\degree$ at 100~MeV to 0.15$\degree$ at 10~GeV\footnote{The
    size of the \FermiL PSF depends on which region of the instrument
    the pair conversion occurs in. In this analysis the size of the ROI
    was tied to the size of the ``Front'' PSF.}. In
  order to maintain sufficient statistics at high energies, the radius
  of the ROI was fixed to 3$\times$0.15$\degree$ above 10~GeV.
\item The \FermiL analysis tools \texttt{gtselect}, \texttt{gtmktime},
  \texttt{gtbin} and \texttt{gtexposure} are then run over each pulsar
  with the corresponding radial and energy selections for all
  observations performed within the period of validity of the pulsar
  timing solution.
\item The photon event list is then barycentered and phase-folded
  using the \texttt{Tempo2} package \citep{Hobbs2006MNRAS} with the
  \Fermi \texttt{Tempo2} plugin and the corresponding timing solution.
\item For the 66 pulsars where timing solutions other than those
  provided by the 2PC were used, a cross-correlation analysis was
  performed on the phase data to determine the phase offset between
  the derived pulsar light curve and the light curve published in the
  2PC supplementary material. The phase data is then corrected for
  this phase offset to ensure that the (arbitrary) 2PC definition of
  phase zero is maintained in this analysis so that the defined \Off
  phase region remains valid.
\item Within each energy bin, a cut on phase is applied and events
  which fall within the \Off phase region and those which fall outside
  this region - the \On phase region - are selected. The ratio of the
  size of the \On phase range to the size of the \Off phase range,
  defined as $\alpha$, is then used to scale the number of event
  counts in the \Off phase region (N$_{\rm off}$) to the number in the
  \On region (N$_{\rm on}$).
\item The number of excess pulsed events is then defined as N$_{\rm
  ex}$$=$N$_{\rm on}$$-$$\alpha$N$_{\rm off}$ and the flux is N$_{\rm ex}$
  divided by the exposure ($\mathcal{T}$) calculated in step 3 using
  \texttt{gtexposure}. The significance of the excess is calculated
  using Equation~17 from \cite{LiMa1983ApJ}.
\end{enumerate}
Following this procedure we derive the energy distributions for the
\On and \Off phase regions, and the instrument exposure, for all 115
pulsars under study. These distributions, along with the derived AP
SED, are shown for the Crab pulsar in Figure~\ref{fig:Crab}. The SEDs
for several other pulsars derived using the AP method are plotted in
Figure~\ref{fig:sed}. Using the AP method, 19 pulsars are detected
above 10~GeV and all, bar one, appear in the \fermi-LAT catalog of
sources above 10~GeV \citep{Ackermann2013ApJS}. The properties of
these 19 pulsars are listed in Table~\ref{tab:10GeVsummary}.
\begin{figure*}
\centering 
\includegraphics[width=0.99\textwidth]{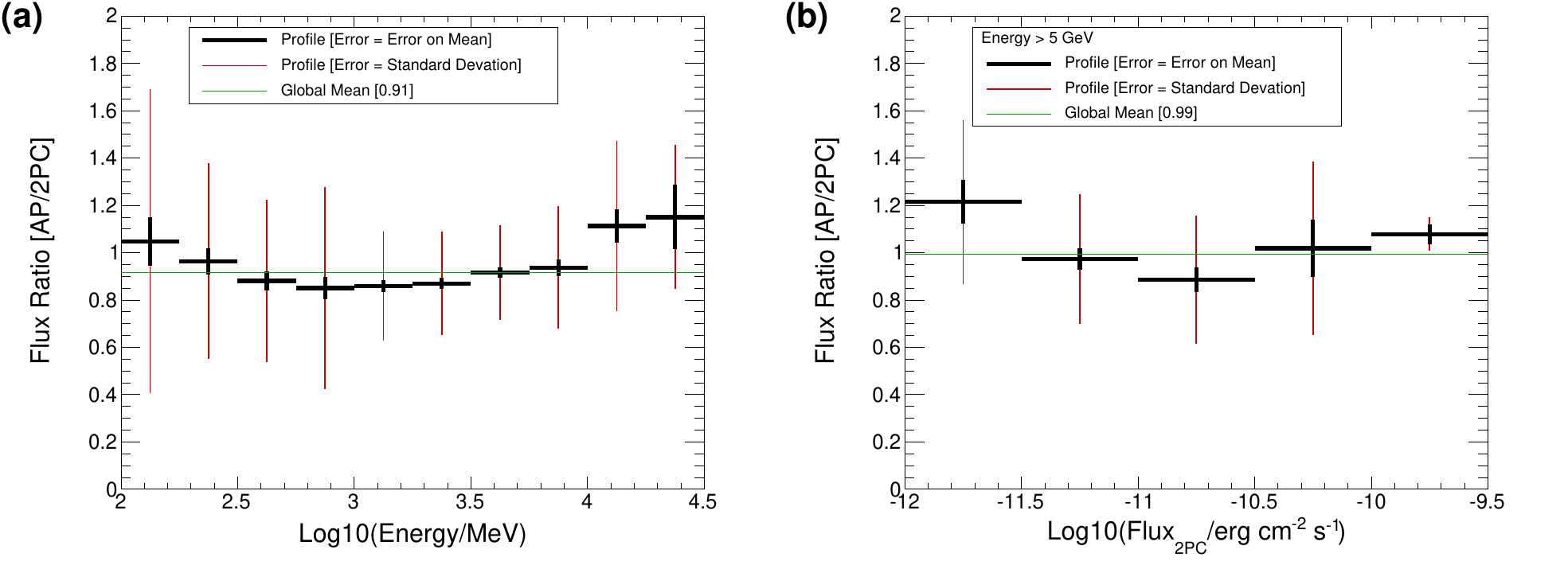}
\caption{\small The ratio of the AP flux to the 2PC flux plotted in profile
  histograms for 83 of the pulsars under study. In panel (a) the ratio
  is plotted against the bin energy while in panel (b) the ratio is
  plotted against the measured 2PC flux for bin energies above 5~GeV.}
\label{fig:Ratio}
\end{figure*}

\begin{table}
\centering
\begin{tabular}{cccc}\hline\hline
Name           &  Excess  & Flux              & Significance \\ 
               &          & [$\times10^{-10}$  &   $\sigma$   \\  
               &          &  cm$^{-2}$s$^{-1}$] &              \\ \hline
PSR J0007+7303   &  289.7$\pm$18.9  & 11.6$\pm$0.8 & 15.4  \\     
PSR J0534+2200   &  377.3$\pm$41.4  & 21.7$\pm$2.4 & 9.1   \\     
PSR J0633+0632   &  11.4$\pm$3.4  & 0.7$\pm$0.2 & 3.4      \\     
PSR J0633+1746   &  542$\pm$52.6  & 32.0$\pm$3.1 & 10.3    \\     
PSR J0835$-$4510 &  1703$\pm$60.4  & 98.1$\pm$3.5 & 28.2   \\     
PSR J1028$-$5819 &  81.8$\pm$12.1  & 4.3$\pm$0.6 & 6.7     \\     
PSR J1048$-$5832 &  32$\pm$9.4  & 1.7$\pm$0.5 & 3.4        \\     
PSR J1413$-$6205 &  46.7$\pm$12.3  & 2.4$\pm$0.6 & 3.8     \\     
PSR J1420$-$6048 &  44.3$\pm$13.6  & 2.3$\pm$0.7 & 3.2     \\     
PSR J1620$-$4927 &  43.5$\pm$13.5  & 2.5$\pm$0.8 & 3.2     \\     
PSR J1709$-$4429 &  373$\pm$22.4  & 21.5$\pm$1.3 & 16.6    \\     
PSR J1732$-$3131 &  47.3$\pm$11.8  & 2.8$\pm$0.7 & 4.0     \\     
PSR J1809$-$2332 &  96.3$\pm$12.5  & 5.9$\pm$0.8 & 7.7     \\     
PSR J1907+0602   &  47.1$\pm$9.9  & 2.9$\pm$0.6 & 4.8      \\     
PSR J2017+0603   &  18.1$\pm$3.9  & 1.1$\pm$0.2 & 4.6      \\     
PSR J2021+3651   &  86.8$\pm$14.9  & 4.7$\pm$0.8 & 5.8     \\     
PSR J2032+4127   &  43.2$\pm$7.5  & 2.3$\pm$0.4 & 5.8      \\     
PSR J2111+4606   &  23.8$\pm$5.3  & 1.2$\pm$0.3 & 4.5      \\     
PSR J2229+6114   &  53.6$\pm$9.4  & 2.3$\pm$0.4 & 5.7      \\\hline
\end{tabular}
\caption{\small Properties of 19 pulsars detected in the 10-17~GeV
  energy bin with the AP method whose excess exceed 3$\sigma$. The
  Crab pulsar is excluded from the stacking analysis but is included
  here for completeness. The excesses, fluxes and significances are
  quoted for the 10-17~GeV energy range only. All entries, bar
  PSR~J1732$-$3131, are also listed in the 1FHL
  catalog. PSR~J2017+0603 is the only millisecond pulsar in the list.}
\label{tab:10GeVsummary}
\end{table}

\subsection{Performance of The Aperture Photometry Method}\label{sec:APPerf}
Assuming that the 2PC flux values are indeed more accurately measured
than the AP flux, we can examine the ratio of the AP flux to the 2PC
flux to assess the performance of the AP method. This ratio is plotted
for individual pulsars below the SED panels in Figures~\ref{fig:Crab}
and \ref{fig:sed}. Profile histograms of this ratio for 83 of the
pulsars under study as shown in Figure~\ref{fig:Ratio}. 33 pulsars are
excluded from these profiles and account for cases where no SED is
derived in the 2PC (10), or where no $>$1.5$\sigma$ excess are derived
in the AP analysis (3), or where the pulsar is so weak in the 2PC
analysis that a 2 bins-per-decade binning is chosen instead of the
usual 4 (20).

Examining Figure~\ref{fig:Ratio}a we see that on average, the AP flux
is $\sim$10\% lower than the 2PC flux, although there are trends
affecting the ratio in different energy ranges. Above 10~GeV the AP
flux is, on average, over-estimated by $\sim$15-20\%. There is also a
sizable scatter, particularly at low energies (E$<$500~MeV) were the
r.m.s.~of the flux ratio can be as large as 60\%. It is important to
point out that this AP framework does not measure the total flux from
a pulsar but the difference in the flux level between the \On and \Off
phase regions. In contrast, the likelihood fitting employed in the
\fermi-LAT analysis measures the total flux over all phases, except in
(8) cases like the Crab pulsar where the \Off phase region is used to
model un-pulsed nebular-like emission which is then subtracted from
the total. Given that this AP method does not measure the total flux
over all phases, it is not surprising that it returns a lower flux
than the 2PC, on average. Figure~\ref{fig:Ratio}b shows that above
5~GeV, the average flux ratio is $\sim$1 with only a weak dependence
on the flux value, although at low fluxes (Flux$_{\rm
  2PC}<$3$\times$10$^{-12}\,\,{\rm ergs}\,\,{\rm s}^{-1}{\rm
  cm}^{-2}$) the AP flux is, on average, 20\% higher with a 30\%
scatter. Given this behavior of the flux ratio at high energies and
low fluxes, we estimate a systematic uncertainty at the level of 60\% on
the stacked flux values and limits presented above 30~GeV in
Section~\ref{sec:Res}.

We wish to point out that these systematic uncertainties do not effect
our ability to detect emission, only our ability to derive the
corresponding flux level or flux limit. Since the SED is expected to
fall exponentially above the GeV break in curvature radiation models,
any significant stacked detection above $\sim$100~GeV would lie in
strong tension with exponential cutoff predictions, regardless of the
uncertainty on the corresponding flux level.

\subsection{Stacking Analysis}\label{sec:Stacking}
Knowing the $\alpha$ values, and having calculated the distribution of
N$_{\rm on}$, N$_{\rm off}$ and $\mathcal{T}$ versus energy for each
pulsar, it is quite simple to determine the total excess,
\begin{equation}
{\rm Ex}_{\rm tot}^{b} = \sum_{i=1}^{\rm N}({\rm N}_{\rm on}^{i,b} - \alpha^{i}{\rm N}_{\rm off}^{i,b})
\end{equation}
the total exposure,
\begin{equation}
\mathcal{T}_{\rm tot}^{b} = \sum_{i=1}^{\rm N}\mathcal{T}^{i,b}
\end{equation}
and thus, the average flux,
\begin{equation}
{\rm Flux}_{\rm av}^{b} = \frac{{\rm Ex}_{\rm tot}^{b}}{\mathcal{T}_{\rm tot}^{b}}
\end{equation}
for N pulsars in a given energy bin, $b$. The corresponding
significance of the total excess is determined by the generalized
version of Equation~17 from \cite{LiMa1983ApJ} (see
\citealt{Aharonian2004A&A}), which accepts subsets of data with
different $\alpha$ values. For cases where the significance of the
total excess is less than 2$\sigma$, the method of \cite{Helene1983}
is used to derive the 95\% confidence-level upper limit on the
total excess, which is in turn divided by the total exposure to
determine the corresponding flux upper limit. These procedures are
used to derived the results presented in the following section.

\begin{figure*}
\centering 
\textbf{All Pulsars Stacked}\par\medskip
\includegraphics[width=0.99\textwidth]{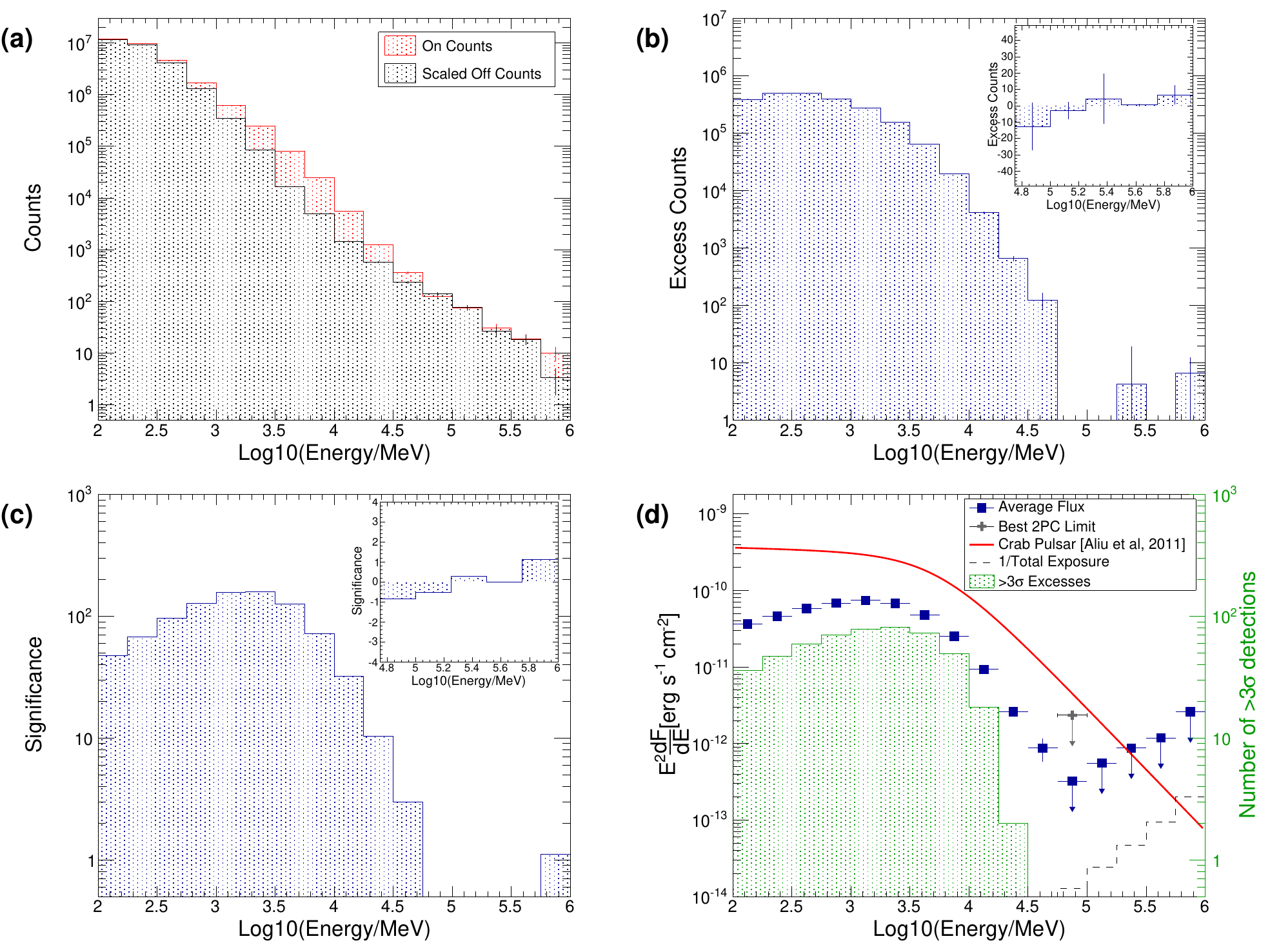}
\caption{\small Stacked analysis for all 115 pulsars under study. Panel (a)
  shows the energy distribution of the total event counts from all \On
  phase regions alongside the total $\alpha$-scaled event counts from
  all \Off phase regions. Panels (b) and (c) show the corresponding
  excesses and significances, respectively. The inserts in these
  panels show the same distributions with a zoom-in on the bins above
  $\sim$50~GeV and with a linear scale on the y-axis. Panel (d) shows
  the average flux (square markers) determined by dividing the total
  excess by the total exposure (see Section~\ref{sec:Stacking}).  The
  dashed-line histogram shows one over the total exposure, indicating
  the flux which would correspond to a single excess photon. This is
  the minimum possible flux which could be measured given the total
  exposure. The gray cross shows the most constraining limit on
  emission from a single pulsar in the 56.2$-$100~GeV range presented
  in the 2PC. The 2PC presented no limits at higher energies. The
  broken power-law fit to the Crab pulsar data from \cite{Aliu2011Sci}
  is plotted for scale. The dot-filled histogram (y-axis scale is on
  the right) indicates how many individual pulsars showed excesses
  which measured above 3$\sigma$ in each energy range. 18 were
  detected in the 10-17~GeV energy bin and they a listed in
  Table~\ref{tab:10GeVsummary}.}
\label{fig:PATot}
\end{figure*}

\section{Results}\label{sec:Res}
\begin{table}
\centering
\begin{tabular}{cccccc}\hline\hline
\vspace{-2mm}\\\vspace{1mm}
Name &  $L_{\gamma}$ & $\dot{E}$ &  $\dot{E}/d^{2}$ & $F_{100}$ & $F^{\rm nt}_{X}$ \\ \hline
PSR J0205+6449 &  -  & $\bullet$  &  $\bullet$  & -   &  $\bullet$  \\
PSR J0437$-$4715 & -   & -   &  -  &   -       &  $\bullet$     \\
PSR J0633+1746   &  -  &  -  & $\bullet$   &  $\bullet$ &  -   \\
PSR J0835$-$4510 &  -  & $\bullet$  & $\bullet$   &  $\bullet$ &  $\bullet$ \\
PSR J1023$-$5746 &  -  & $\bullet$  &  - &  -  & -\\
PSR J1048$-$5832 & $\bullet$  &  -  &  -  &  -  & - \\
PSR J1112$-$6103 & $\bullet$  & -   &  -  & -   &  - \\
PSR J1119$-$6127 & $\bullet$  &  -  & -   & -   & -    \\
PSR J1124$-$5916 &  -  & $\bullet$  &  $\bullet$  &  -  &   $\bullet$   \\
PSR J1410$-$6132 & $\bullet$  & $\bullet$  & -   &  -  &  -   \\
PSR J1418$-$6058 &  -  &  -  &  $\bullet$ &  $\bullet$ &  -   \\
PSR J1420$-$6048 & $\bullet$  & $\bullet$  &  -  &  -  &  -   \\
PSR J1513$-$5908 &  -  & $\bullet$  &  $\bullet$  & -   &  $\bullet$   \\
PSR J1709$-$4429 & $\bullet$  &  -  & $\bullet$  & $\bullet$  &  -   \\
PSR J1747$-$2958 & $\bullet$  &  -  &   - &   - &  $\bullet$   \\
PSR J1801$-$2451 &  -  &  -   &  -   &  -   &  $\bullet$     \\
PSR J1809$-$2332 &  -  &  -  &  -  & $\bullet$  &   -  \\
PSR J1813$-$1246 &  -  & $\bullet$  &  - & $\bullet$  &  -   \\
PSR J1826$-$1256 &  -  &  -  &  - & $\bullet$  &  -  \\
PSR J1833$-$1034 &  -  & $\bullet$  &  $\bullet$  &  -  & $\bullet$   \\
PSR J1836+5925 &  -  &  -  &  -  &  $\bullet$ &   -  \\
PSR J1907+0602 & $\bullet$  &  -  & -   &  -  &  -   \\
PSR J1952+3252 &  -  &  -  &  $\bullet$  & -  &  $\bullet$    \\
PSR J2021+3651 & $\bullet$  & -   & -   &  $\bullet$ &   -  \\
PSR J2021+4026 & $\bullet$  & -   & -   &  $\bullet$ &   -  \\
PSR J2229+6114 &  -  & $\bullet$  &  $\bullet$  &  -  &   $\bullet$  \\\hline
\end{tabular}
\caption{The list of the pulsars used in the sub-sample stacking, where
  the $\bullet$ symbol denotes the inclusion of the pulsar in the
  given sample.}\label{tab:subsamp}
\end{table}

\begin{figure*}
\centering 
\textbf{All Young Pulsars Stacked}\par\medskip
\includegraphics[width=0.99\textwidth]{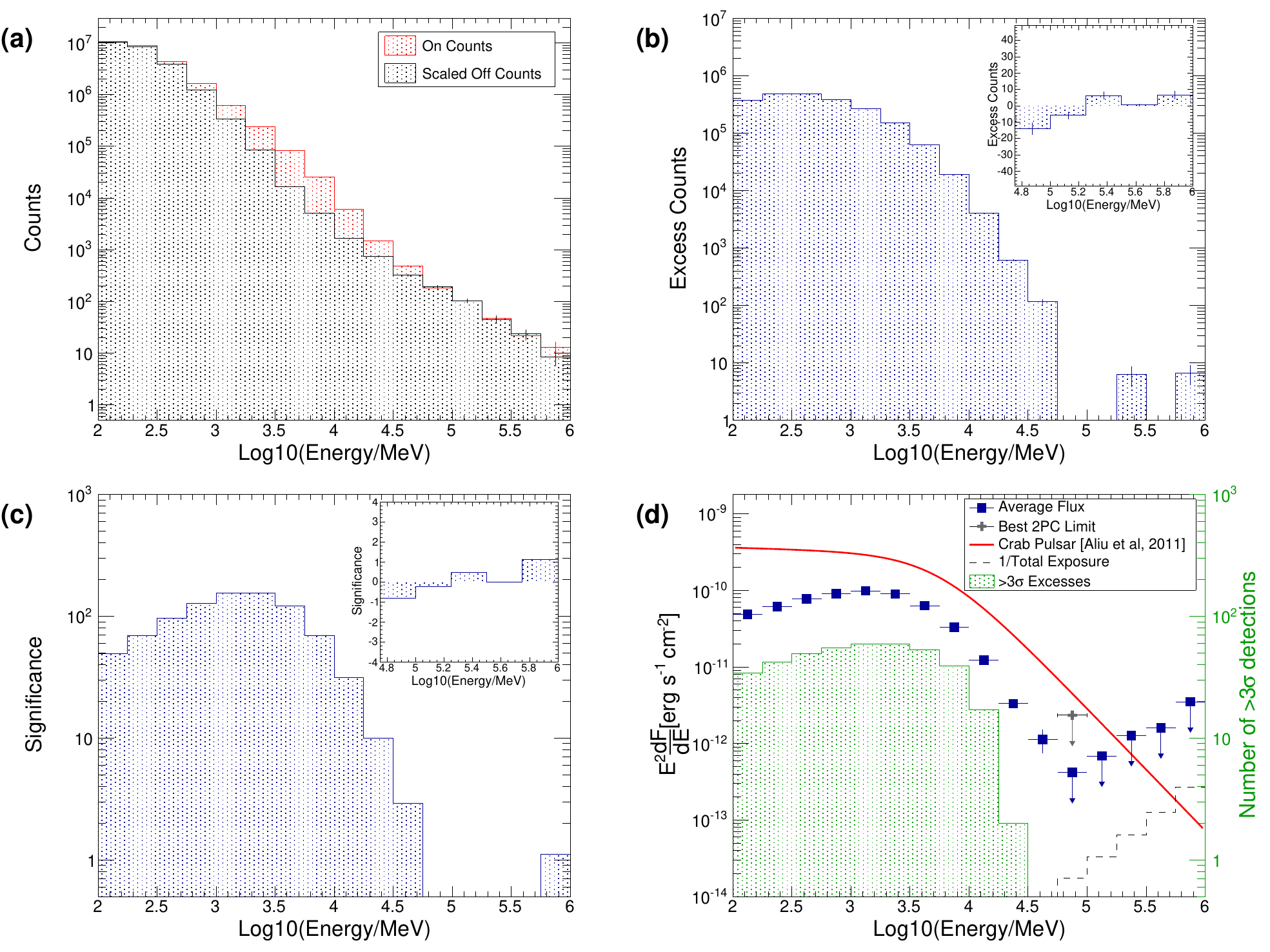}
\caption{\small Stacked analysis for all 76 young pulsars under study. See
  the caption of Figure~\ref{fig:PATot} for a full description of each
  panel.}
\label{fig:PATotyp}
\end{figure*}

\begin{figure*}
\centering 
\textbf{All Millisecond Pulsars Stacked}\par\medskip
\includegraphics[width=0.99\textwidth]{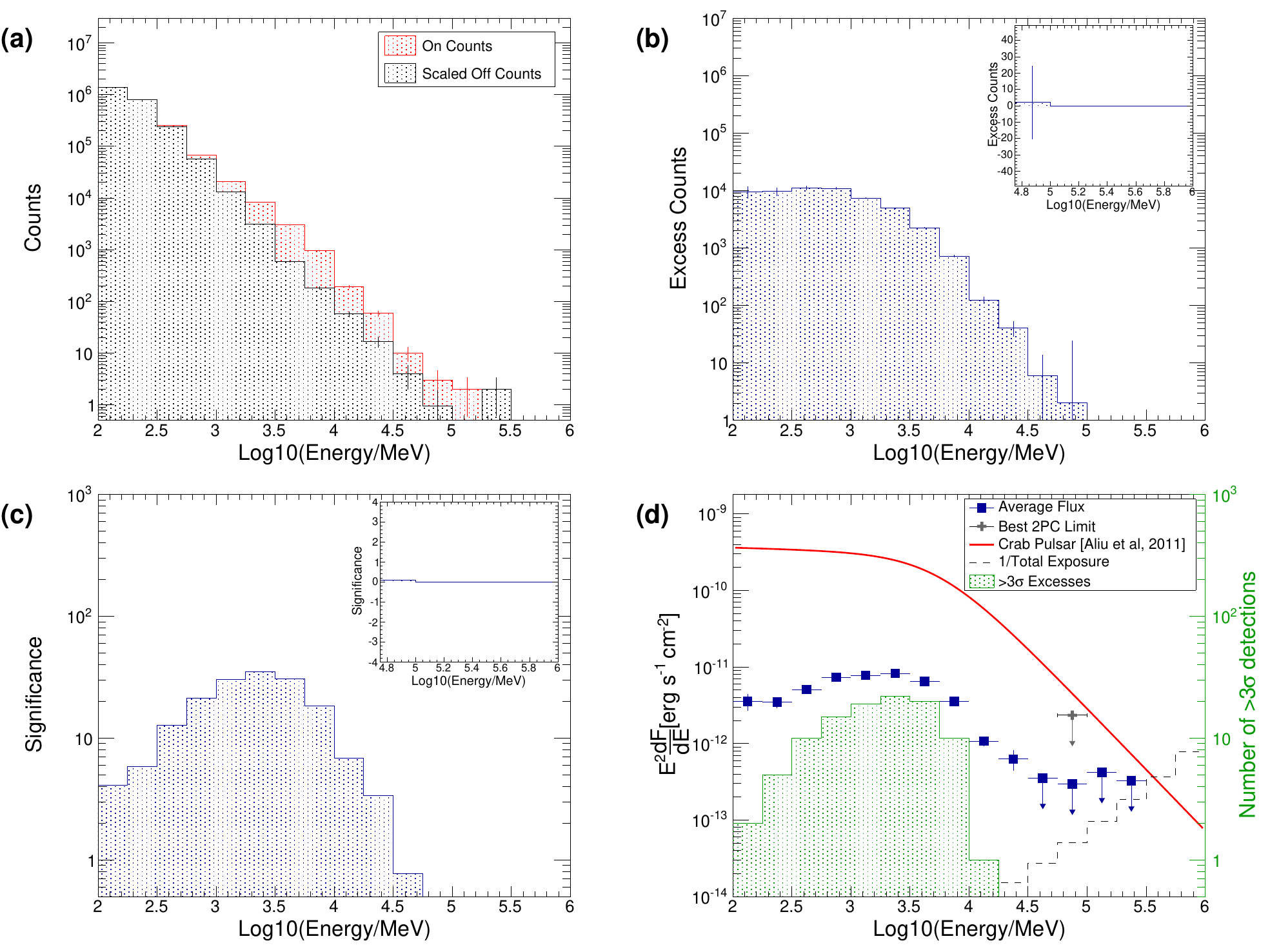}
\caption{\small Stacked analysis for all 39 millisecond pulsars under
  study. See the caption of Figure~\ref{fig:PATot} for a full
  description of each panel. PSR~J2017+0603 is the only millisecond
  pulsar with a $>$3$\sigma$ excess measured in the 10-17~GeV energy
  bin (see Table~\ref{tab:10GeVsummary} for a description of this
  excess).}
\label{fig:PATotmsp}
\end{figure*}

The results of the stacking analysis for the entire pulsar sample are
presented in Figure~\ref{fig:PATot}. The stacking analysis results for
the separate young pulsar and millisecond pulsar ensembles are shown
in Figures~\ref{fig:PATotyp} and ~\ref{fig:PATotmsp}, respectively. No
significant excesses are seen in any of these analyses at energies
above 50~GeV. Upper limits on the average flux, determined at the 95\%
confidence-level, are listed in Table~\ref{tab:Stacksummary} for three
energy bins above 50~GeV. Limits are also presented in units of the
Crab pulsar where the broken power-law fit to the \FermiL and VERITAS
data presented in \cite{Aliu2011Sci} defines a Crab pulsar unit.

In addition to these analyses, we stacked sub-samples of the data
where each sub-sample was composed of the 10 pulsars with the largest
value of a given parameter. Sub-sample selections based on gamma-ray
luminosity ($L_{\gamma}$), spin-down power ($\dot{E}$), spin-down
power over distance squared ($\dot{E}/d^{2}$), gamma-ray photon flux
($F_{100}$) and non-thermal X-ray energy flux ($F^{\rm nt}_{X}$) were
investigated and are listed in Table~\ref{tab:subsamp}\footnote{The
  Crab pulsar was excluded from all of these sub-sample stacking
  analyses. The parameter values listed in the 2PC catalog were used
  in all cases.}. No significant excesses were observed above 50~GeV
in any of these sub-sample stacking analyses.

\begin{figure}
\centering 
\includegraphics[width=0.49\textwidth]{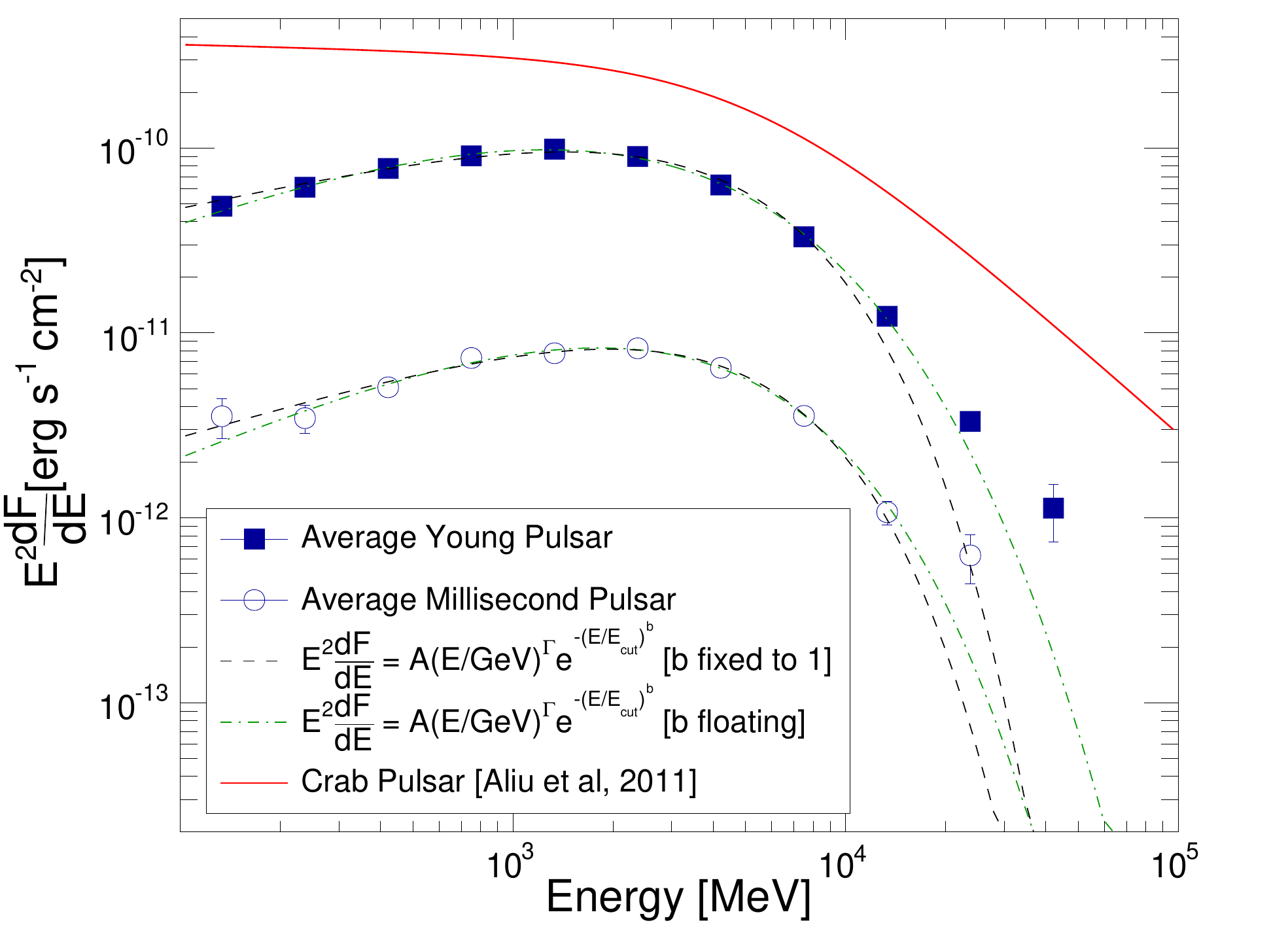}
\caption{\small The average SEDs derived from the stacking of the 76 young
  pulsars and 39 millisecond pulsars. The SEDs are each fit with a
  power law times a super-exponential cutoff keeping $b$ both fixed to
  unity and allowing it to float. For the pure exponential cutoff case
  ($b=1$) the best fit $\Gamma$ value is 0.54$\pm$0.05 for the
  millisecond pulsars and 0.41$\pm$0.01 for the young pulsars while
  the best fit $E_{\rm cut}$ values are 3.60$\pm$0.21~GeV and
  3.54$\pm$0.04~GeV, respectively. Allowing $b$ to float we find that
  sub-exponential forms ($b<1$) are preferred, with the best-fit $b$
  value of 0.59$\pm$0.02 for the young pulsars and 0.7$\pm$0.15 for
  the millisecond pulsars. The broken power-law fit to the Crab pulsar
  data from \cite{Aliu2011Sci} is plotted for scale.}
\label{fig:YPVMSP}
\end{figure}

The shape of the average young pulsar and average millisecond pulsar
SEDs were categorized by fitting a power law times a super-exponential
cutoff function
\begin{linenomath}
\begin{equation}
E^{2}\frac{dF}{dE} = A {\left(\frac{E}{\rm 1~GeV}\right)}^{\Gamma}e^{-\left(\frac{E}{E_{\rm cut}}\right)^{b}}
\end{equation}
\end{linenomath}
to the SED data. These fits are presented in
Figure~\ref{fig:YPVMSP}. Fixing $b=1$ reduces Equation~4 to a power
law times an exponential cutoff function and, as expected, this
functional form does not reproduce the sub-exponential fall of the SED
above the break. However it can be used to measure the average
flux-weighted value of the spectral index ($\Gamma$) and cutoff
($E_{\rm cut}$) parameters \citep{Abdo2013ApJS}. It is clear from
Figure~\ref{fig:YPVMSP} that the average SEDs have qualitatively the
same shape, with the average flux from the 39 millisecond pulsars
about an order of magnitude lower than the average flux from the 76
young pulsars. The spectral parameters derived from the fitting are
both remarkably similar. The best fit $\Gamma$ value is 0.54$\pm$0.05
for the millisecond pulsars and 0.41$\pm$0.01 for the young pulsars
while the best fit $E_{\rm cut}$ values are 3.60$\pm$0.21~GeV and
3.54$\pm$0.04~GeV, respectively. Allowing $b$ to float we find that
sub-exponential forms ($b<1$) are preferred, with the best-fit $b$
value of 0.59$\pm$0.02 for the young pulsars and 0.7$\pm$0.15 for the
millisecond pulsars. Note that only statistical uncertainties on the
SED data points were used during the fitting and thus the uncertainty
on the best-fit parameter values are likely underestimated.

\begin{table*}
\centering
\begin{tabular}{ccc|cc|cc|cc}\hline
      & &                          & \multicolumn{2}{|c|}{All}       & \multicolumn{2}{|c|}{Young Pulsars} & \multicolumn{2}{|c}{Millisecond Pulsars}   \\\hline
\multicolumn{3}{c|}{Energy Range} & Flux Limit       & Flux Limit   & Flux Limit      & Flux Limit    & Flux Limit       & Flux Limit   \\
\multicolumn{3}{c|}{[GeV]}        & [$\times10^{-12}$ & [Crab pulsar & [$\times10^{-12}$ & [Crab pulsar & [$\times10^{-12}$ & [Crab pulsar  \\
\multicolumn{3}{c|}{}             & cm$^{-2}$s$^{-1}$] &  units]      & cm$^{-2}$s$^{-1}$] &  units]      & cm$^{-2}$s$^{-1}$] &  units]      \\\hline
 56.2 &---& 100                        &  1.57            & 0.07         &  2.03            & 0.09         &  1.44            & 0.07         \\
  100 &---& 177                        &  1.52            & 0.31         &  1.88            & 0.38         &  1.14            & 0.23         \\ 
  177 &---& 316                        &  1.34            & 1.21         &  1.96            & 1.76         &  0.50            & 0.45         \\\hline
\end{tabular}
\caption{\small Limits at the 95\% confidence level on the average
  flux from stacked ensembles of gamma-ray pulsars. The limit values
  presented in Crab pulsar units assume the broken power-law fit to
  the Crab pulsar data from \cite{Aliu2011Sci} is a Crab pulsar flux
  unit.}
\label{tab:Stacksummary}
\end{table*}

\section{Discussion and Conclusion}\label{sec:Conclusion}
Following a stacked analysis of 115 gamma-ray pulsars, with an average
exposure of $\sim$4.2~yr per pulsar, we find no evidence of cumulative
emission above 50~GeV. Stacked searches exclusive to the young
pulsars, the millisecond pulsars, and several other promising
sub-samples also return no significant excesses above 50~GeV. Any
average emission present in the entire pulsar sample is limited to be
below $\sim$7\% of the Crab pulsar in the 56-100~GeV band and to be
below $\sim$30\% in the 100-177~GeV band. The average flux limits
presented in Table~\ref{tab:Stacksummary} are roughly 3 times lower
than the best flux limits achieved in dedicated individual pulsar
analyses done in the 2PC in the 56-100~GeV band.

One should note that a limit on the average flux from 115 pulsars at
7\% of the Crab pulsar level is consistent with, for example, a
scenario in which all 115 pulsars emit at 7\% of the Crab pulsar
level. It is also consistent with a scenario in which 8 pulsars emit
at 100\% the level of the Crab pulsar and the remaining 107 pulsars
have zero emission. Therefore this analysis does not exclude the
possibility of finding several pulsars which are as bright as the Crab
pulsar above 50 GeV, or several dozen which are ten times
dimmer \footnote{This point is illustrated by the fact that the Vela
  pulsar has recently been shown to emit at $\sim$130\% of the Crab
  pulsar level in the 50$-$100~GeV energy range
  \citep{Leung2014}.}. It does, however, constrain the average flux
from the ensemble, and therefore for every individual pulsar detected
above this flux limit, the average emission from the remaining pulsars
is constrained to be further below the limit.

In the 100~MeV to $\sim$50~GeV energy range we find that the average
SEDs returned from the young pulsar and millisecond pulsar stacking
analyses are very similar in shape and are generally compatible with a
power law times a sub-exponential cutoff. \cite{Abdo2010ApJ} and
\cite{Celik2011AIPC} have shown that a sub-exponential cutoff function
approximates a superposition of exponential cutoffs, thus the
appearance of a sub-exponential cutoff in the ensemble SED is to be
expected within a curvature radiation model. We note, however, that
the highest energy spectral point is higher than the best fit
sub-exponential cutoff function at the $\sim$2.4$\sigma$ level in both
the young pulsar and millisecond pulsar cases. This cannot be taken as
strong evidence for a non-exponentially-suppressed pulsar emission
component aggregating in the stacked analysis, however, the available
data cannot rule it out beyond the level of the limits shown in
Figures~\ref{fig:PATot},~\ref{fig:PATotyp} and \ref{fig:PATotmsp} and
Table~\ref{tab:Stacksummary}.

At energies above 100~GeV, individual pulsar limits made by air
Cherenkov telescopes are much stronger than those achievable with the
\fermiL. The sensitivity necessary to detect the Crab pulsar emission
above 100~GeV at the $\sim$5$\sigma$ level is achieved by VERITAS in
under 30~hrs, for example. Beyond this work, improvements can be made
using the forthcoming \FermiL pass-8 data release which will improve
the \fermi-LAT acceptance by $\sim$25\% at 100~GeV
\citep{Atwood2013arXiv}. Improvements to this stacking analysis can
also be made by employing a likelihood framework to stack the sources
(see \citealt{Ackermann2011PhRvl} for example), rather than the simple
\On minus \Off procedure described here. The flux sensitivity of any
stacking analysis will, however, ultimately be bounded by the exposure
of the \fermiL. The dashed-line histograms (one over the total
exposure) in Figures~\ref{fig:PATot},~\ref{fig:PATotyp} and
\ref{fig:PATotmsp} indicate that limits derived in this analysis are
factors of $\sim$4$-$20 times larger than the minimum measurable
average flux in the 100$-$177 GeV range. A future stacking analysis
which doubles both the number of pulsars and the duration of
observation used will increase the exposure term by factor of 4,
indicating that future stacking analyses which do not yeild detections
may improve on the limits presented here by perhaps one or two orders
of magnitude.
 
\acknowledgments The author is supported in part by the Kavli
Institute for Cosmological Physics at the University of Chicago
through grant NSF PHY-1125897 and an endowment from the Kavli
Foundation and its founder Fred Kavli. I am grateful to Pat Moriarty,
David Hanna, Nepomuk Otte and Benjamin Zitzer for their comments on
the early drafts of this manuscript. I appreciate the anonymous
referee for the comments and suggestions made during the review
process.

\end{document}